\documentclass[twocolumn,3p,times,procedia]{elsarticle}

\usepackage{ecrc}
\usepackage{graphicx}
\usepackage{balance}
\usepackage{array,threeparttable}
\usepackage{url}

\usepackage{endnotes} 
\usepackage{amsmath}
\usepackage{enumitem}
\usepackage{float}
\usepackage{caption}
\usepackage{subcaption}
\usepackage{graphicx}  
\usepackage{xcolor}
\usepackage{color}
\usepackage[table]{xcolor}

\usepackage{titlesec}

\titleformat{\paragraph}
  {\normalfont}          
  {\theparagraph}        
  {1em}                  
  {}             

\volume{00}

\firstpage{1}

\runauth{}

\jid{procs}

\CopyrightLine{202X}{Published by Elsevier Ltd.}

\usepackage{amssymb}
\usepackage{amsmath}
\usepackage{multicol}
\usepackage{algorithm}
\usepackage{algpseudocode}

\usepackage[figuresright]{rotating}
\usepackage{bm}
\usepackage{caption}
\usepackage{titlesec}
\titleformat*{\section}{\small \bf}
\titleformat*{\subsection}{\small \em}
\titleformat*{\subsubsection}{\small \em}

\usepackage{fancyhdr}
\fancyheadoffset[RO,EL]{0pt}
\usepackage{geometry}
\begin{document}\small
\begin{frontmatter}




\dochead{}
\title{
\begin{flushleft}
{\LARGE Information and communications technologies for carbon sinks from economics and engineering perspectives} 
\end{flushleft}
}
 %

\author[]{ \leftline {Yuze Dong$^1$$^a$, Jinsong Wu $^*$$^a$$^b$}}

\address{ \leftline {$^a$School of Artificial Intelligence, Guilin University of Electronic Technology, }
  \leftline{Guilin, 541004, China}

  \leftline {$^b$Department of Electrical Engineering, University of Chile, Santiago, 8370451, Chile}

}


\cortext[*]{Corresponding author: Jinsong Wu, Email address: wujs@ieee.org.}


\fntext[1]{Yuze Dong, Email address:dyz459dd@gmail.com.}

\begin{abstract}

Climate change has intensified the urgency of effective carbon sink solutions, yet the integration of Information and Communications Technologies (ICT) in these systems remains fragmented despite its transformative potential. This paper provides a comprehensive analysis of  ICT applications in carbon sink projects from both economic and engineering perspectives, a dual lens approach rarely explored in the existing literature. In carbon trading, blockchain has improved transaction speed by 40\%, while AI-based optimizations have reduced operational costs by 15\% in projects such as Petra Nova. Through systematic examination, we identify three key findings: (1) ICT transforms carbon economics through digital financing platforms and blockchain-based trading systems, with AI enhancing price prediction, though data interoperability remains challenging; (2) digital technologies advance both natural and artificial sequestration from forest monitoring to Carbon Capture, Use and Storage (CCUS) optimization, yet lack integrated real-time control solutions; (3) realizing ICT's full potential requires addressing its environmental costs, strengthening policy support, and fostering interdisciplinary collaboration. By bridging the economic engineering divide and mapping current applications alongside future opportunities, this paper demonstrates that deeper integration of digital technologies is essential to scale carbon sink solutions to meet climate targets.

\end{abstract}


\end{frontmatter}


\section{Introduction}

Human activities are causing unprecedented global environmental changes. Between 2011 and 2020, global surface temperature increased by $1.09^\circ C$ compared to levels of 1850-1900 \cite{lee2023ipcc}, with $1.07^\circ C$ attributed to human actions \cite{lee2023ipcc}. Greenhouse gas emissions from industrialization, primarily $CO_2$ and methane, are the main drivers of warming, with $CO_2$ concentrations reaching record highs by 2019 \cite{lee2023ipcc, ak2021wmo}. Despite improved energy efficiency and carbon intensity, emissions from the energy, industry, transport and construction sectors continue to increase \cite{lee2023ipcc}.

Climate change impacts health and economies by reducing crop yields, increasing sea levels, and increasing coastal protection costs. A warming scenario of $2.5^\circ C$ could cut the average GDP output by 0.7\% \cite{tol2009economic}. It also raises extreme weather risks and infectious diseases such as malaria, which cause 1–2 million child deaths annually \cite{haines2004health}.

$CO_2$ emissions are the main driver of climate change, and net carbon emissions can be expressed as
\[
\text{net carbon emissions} = \text{gross carbon emissions} - \text{carbon sequestration}
\]
This identity suggests two primary mitigation strategies: reducing gross emissions through low-carbon energy transitions or increasing carbon sequestration using natural or engineered methods. The carbon sink projects mentioned in this article belong to the latter.

The development of carbon sink projects relies on cooperation across sectors, from ecology to finance and engineering. In practice, Information and Communication Technologies (ICT) have shown a clear potential to enhance these efforts. For example, blockchain improved transaction speed and reduced costs in carbon trading, while Artificial Intelligence (AI) applications in sequestration projects helped reduce energy use and improve operational safety. These cases suggest that ICT can effectively support both the economic and technical dimensions of carbon sink initiatives. These benefits stem from the fundamental capabilities of ICT, which we briefly introduce below to establish a clearer understanding of its relevance.

ICT refer to the broad spectrum of technologies designed to create, store, process, transmit, and utilize digital information. They encompass hardware (such as computers, sensors, networking devices), software (such as data analytics platforms, AI algorithms), and communication infrastructures (such as the 5-th generation wireless communications (5G), Internet of Things (IoT), cloud computing). By integrating these components, ICT enable real-time data acquisitions, secure exchange,s and intelligent decision-making in diverse fields, from telecommunications and finance to environmental monitoring and smart infrastructure. Its adaptability and scalability make it a fundamental driver of modern digital ecosystems.

In recent years, ICT have been widely applied in green applications \cite{lukose2016review, batool2019green, 10214579}, including areas such as smart energy grids, sustainable infrastructure management, and precision agriculture. However, when they come to carbon sink projects, their integrations remain limited, fragmented, and frequently overlooked. Compared with their applications in other sectors, the intersections between ICT and carbon sinks still lack systematic explorations, practical frameworks, and sustained academic attention.

Against this backdrop, it becomes increasingly valuable to examine how ICT can be systematically integrated into carbon sink initiatives. Although existing efforts offer promising signals,  comprehensive understandings of their applications, challenges, and long-term potential remains limited. This paper aims to contribute to this growing area by reviewing current practices, identifying key technological opportunities and constraints, and outlining possible directions for future development. Through this work, we hope to support a more coordinated and effective use of ICT in both carbon sequestration and carbon finance systems.

In summary, our main contributions are as follows.

\begin{enumerate}
    \item We introduce the operational and engineering foundations of carbon sink projects to help readers quickly grasp their core framework.
    \item We analyze the current integration of ICT in both the economic and engineering aspects of carbon sink systems and offer forward-looking insights into its future applications.
    \item We highlight the potential of ICT to improve efficiency while also addressing its environmental costs, calling for strategies to mitigate associated risks.
\end{enumerate}

To support these contributions, the remainder of the paper is structured into three main parts. The first focuses on ICT applications in the economic operations of carbon sink projects, including financing models, trading systems, and forecasting tools, along with their current limitations. The second part turns to ICT's role in engineering implementation, examining how digital technologies are used in natural and artificial carbon sequestration, and identifying technical bottlenecks. The final part concludes with reflections on environmental impacts, policy implications, and opportunities for future interdisciplinary collaboration.
\section{Carbon sink and carbon sink projects}
\subsection{Carbon sink}
\begin{figure}[H]
    \centering
    \includegraphics[width = 0.43\textwidth]{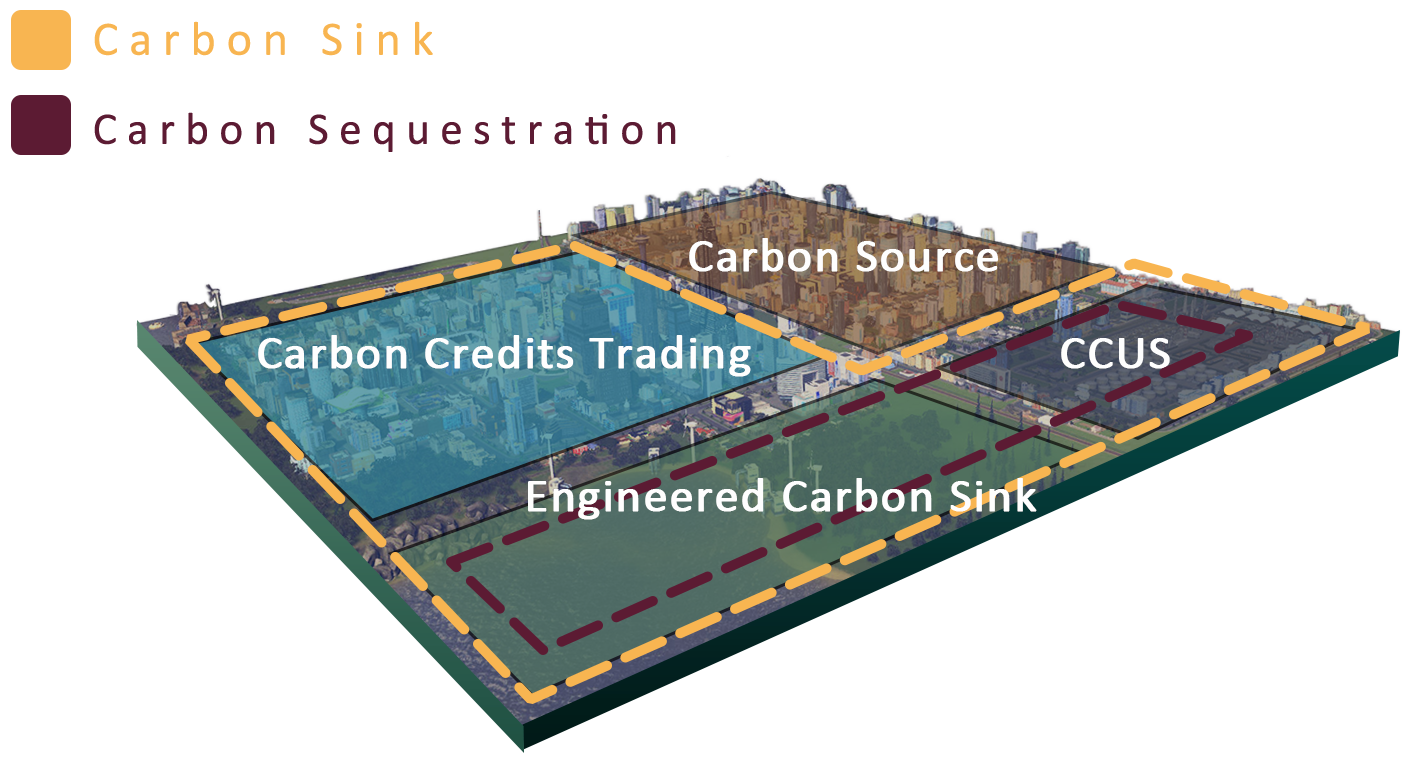}
    \caption{Schematic diagram of carbon sink concepts.}
    \label{fig:carbon_sink_and_source}
\end{figure}
A carbon sink is defined as any process, activity, or mechanism that captures and stores atmospheric carbon dioxide. There are many concepts related to carbon sinks, such as carbon sequestration, carbon sources, and Carbon Capture, Utilization and Storage (CCUS), whose geographical locations are shown in Fig. \ref{fig:carbon_sink_and_source}. This paper provides a comprehensive introduction to the relationships and connections among these concepts.

The concept of a carbon sink was first introduced in the 1992 United Nations Framework Convention on Climate Change (UNFCCC), defining it as any process or activity that removes greenhouse gases from the atmosphere. A carbon source, in contrast, refers to processes that release these gases \cite{UNFCCC1992}. Although the definition of a source is consistent, the definition of a sink varies. The 2005 Intergovernmental Panel on Climate Change (IPCC) described it as the natural absorption of $CO_2$ by soil, forests, or oceans \cite{metz2005ipcc}, while the 2018 IPCC expanded it to include any reservoir, natural or man-made, that stores greenhouse gases \cite{IPCC2018}. Both definitions align with, but emphasize, different aspects of carbon storage.

Carbon sequestration is the act of capturing and storing carbon, often in the form of carbon dioxide, within a designated carbon reservoir \cite{IPCC2018}. Unlike carbon sinks, which cover the entire process of reducing atmospheric greenhouse gases, sequestration focuses specifically on the storage phase. Carbon sinks involve a broader range of activities, including capture, storage, verification, and trading. In this paper, "carbon sequestration" refers to the implementation phase of carbon sink projects.

CCUS has evolved from Carbon Capture and Storage (CCS), which involves capturing $CO_2$ from industrial sources and storage to isolate it from the atmosphere \cite{IPCC2018}. Originally used in the natural gas industry, CCS was later applied to climate change research \cite{metz2005ipcc}. As understanding grew, CCUS emerged, where Carbon Capture and Utilization (CCU) refers to capturing $CO_2$ and using it in products. If $CO_2$ is stored for a long time in the products, known as CCUS \cite{IPCC2018}. CCUS is a critical technology in the carbon sequestration process, particularly during the implementation of the carbon sink project.

\subsection{Carbon sink projects}

A carbon sink project refers to activities that aim to reduce, absorb, or isolate carbon dioxide in the atmosphere. In addition to the commonly seen engineering implementation process, the economic operation model is also an essential component. The overall operational process can be summarized in the following points.

\begin{figure*}
    \centering
    \resizebox{0.7\textwidth}{!}{\includegraphics{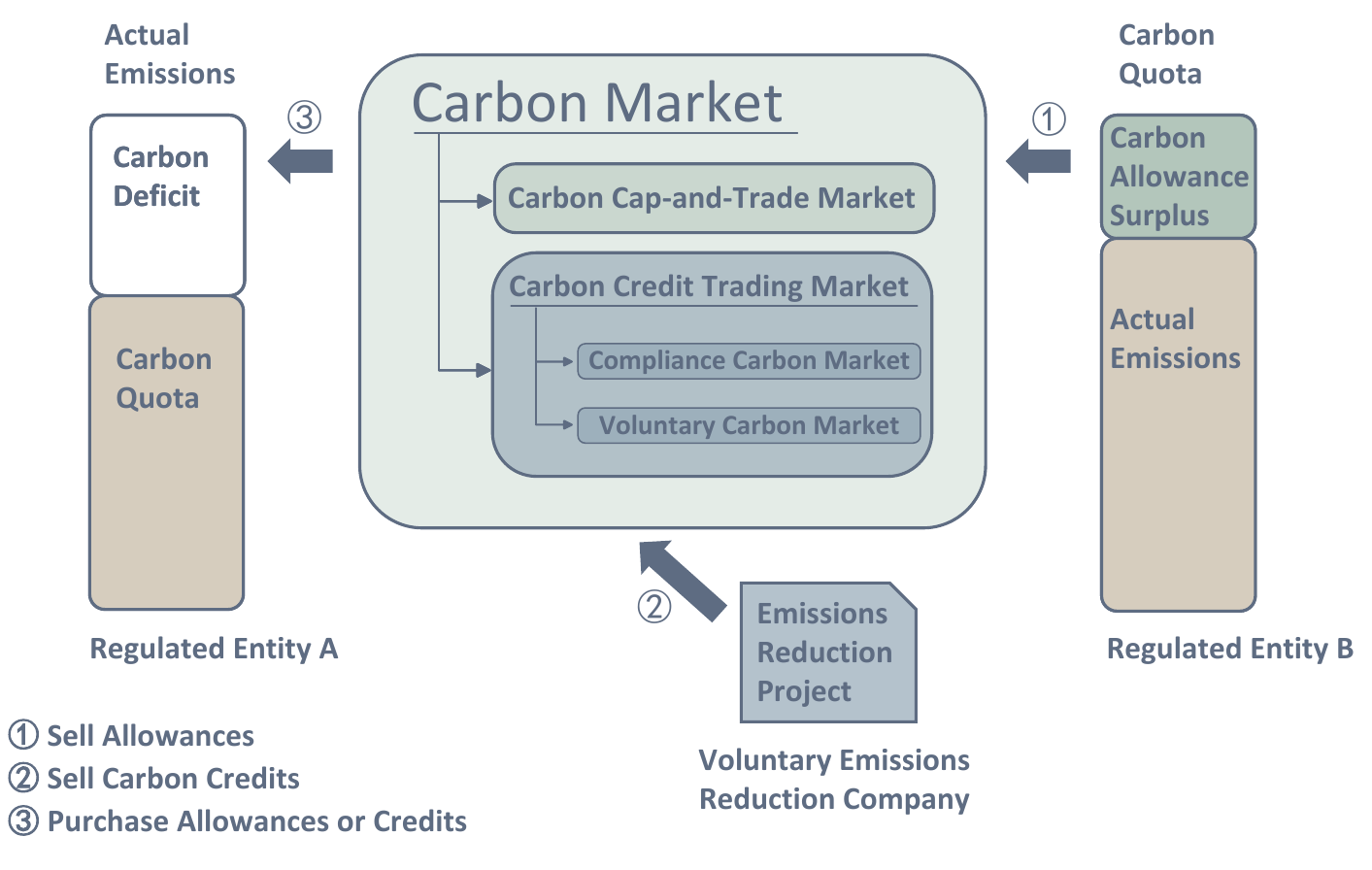}}
    \caption{Structural composition of the carbon market and participating entities.}
    \label{fig:carbon_market}
\end{figure*}

\begin{enumerate}
    \item Implementation phase:
        \begin{enumerate} 
            \item Project preparation includes methodology selection, site planning, carbon estimation, financing, stakeholder participation, and government approval.
            \item The execution of the project covers the implementation and monitoring of engineering to ensure that targets are met.
            \item Third-party verification enables certification and registration for carbon trading.
        \end{enumerate}
    \item Trading phase:
        \begin{enumerate}
            \item Carbon credits are traded under market rules, with offset ratios, credit cancelation, and regulatory oversight ensuring fairness and transparency.
        \end{enumerate}
\end{enumerate}

This paper will comprehensively introduce the current applications of ICT in carbon sink projects from two perspectives: the economic operations and the implementation process, and will provide insight into future prospects.

\section{ICT in carbon sink projects: economic operations}
\subsection{Essential economic concepts for carbon sink projects}
In carbon sink projects, the main economic domains in which information and communication technologies are widely applied are project financing and carbon market trading. According to the World Bank 2023 State and Trends of Carbon Pricing report \cite{weltbank2023state}, 73 carbon pricing instruments were in operation worldwide in 2022, covering approximately 23 percent of global greenhouse gas emissions and generating approximately 95 billion US dollars in revenue. These developments reflect the increasing role of digital technologies such as blockchain and artificial intelligence in enhancing transparency, automating transactions, and supporting carbon finance operations.

\subsubsection{Carbon sink financing}
Financing is crucial for carbon sink projects due to their high initial costs and long payback periods, such as afforestation and CCUS. It supports project initiation, operations, and helps manage risks from market changes, technology, natural events, and policies. Carbon sink financing falls under sustainable finance, which focuses on environmental goals.

Carbon sink projects often use a project financing model. In this model, investors are only responsible for their investment and do not repay debts with personal assets. Returns depend on the project's future cash flows, not the borrower's personal assets, which provides limited liability to investors and borrowers. Typically, a project company, a special purpose vehicle, is created to handle financing based on project assets such as funds, bonds, and credit \cite{AlonsoConde2020OnTE}. This structure limits risks to a single entity, protecting the parent company. It also allows for specialized management of cash flows, costs, and revenues, making it easier to secure financing. For international carbon market projects, an independent project company also streamlines carbon credit generation, verification, and trading.

Currently, carbon sink projects are financed primarily through green funds, green bonds, and green credit, which operate similarly to traditional financing but with more favorable policies. Studies have shown that green funds and bonds outperform traditional options, with Morningstar reporting a 6.9\% return for green funds compared to 6.3\% for traditional ones \cite{bioy2020sustainable}. Green bonds also performed well, particularly in Europe \cite{GIANFRATE2019127}. In 2020, ESG-compliant investment assets reached \$3.1 trillion \cite{us_sif_report}. This financing supports carbon sink projects, boosts green investment, and drives sustainable economic growth.

In addition to financing, risk management plays a crucial role in carbon sink projects, and green insurance is emerging as a widely adopted strategy. Green insurance protects against environmental risks, encompassing both voluntary and mandatory coverage for areas such as carbon credits and emissions trading. For example, China Life Insurance provides "Carbon Sink Insurance" to safeguard against fluctuations in the price of forestry carbon sinks. Natural Based Solutions (NBS) insurance schemes\footnote{ refer to any insurance products and services that directly or indirectly mitigate and manage risks to the natural ecological environment, including disaster cleanup, post-disaster reconstruction and restoration, and property insurance that compensates for business interruptions resulting from natural resource losses.} also contribute to the management of natural risks. Although still in the early stages, these insurance products are under development and are expected to expand in the future.
\begin{figure*}
	\centering
	\resizebox{0.82\textwidth}{!}{\includegraphics{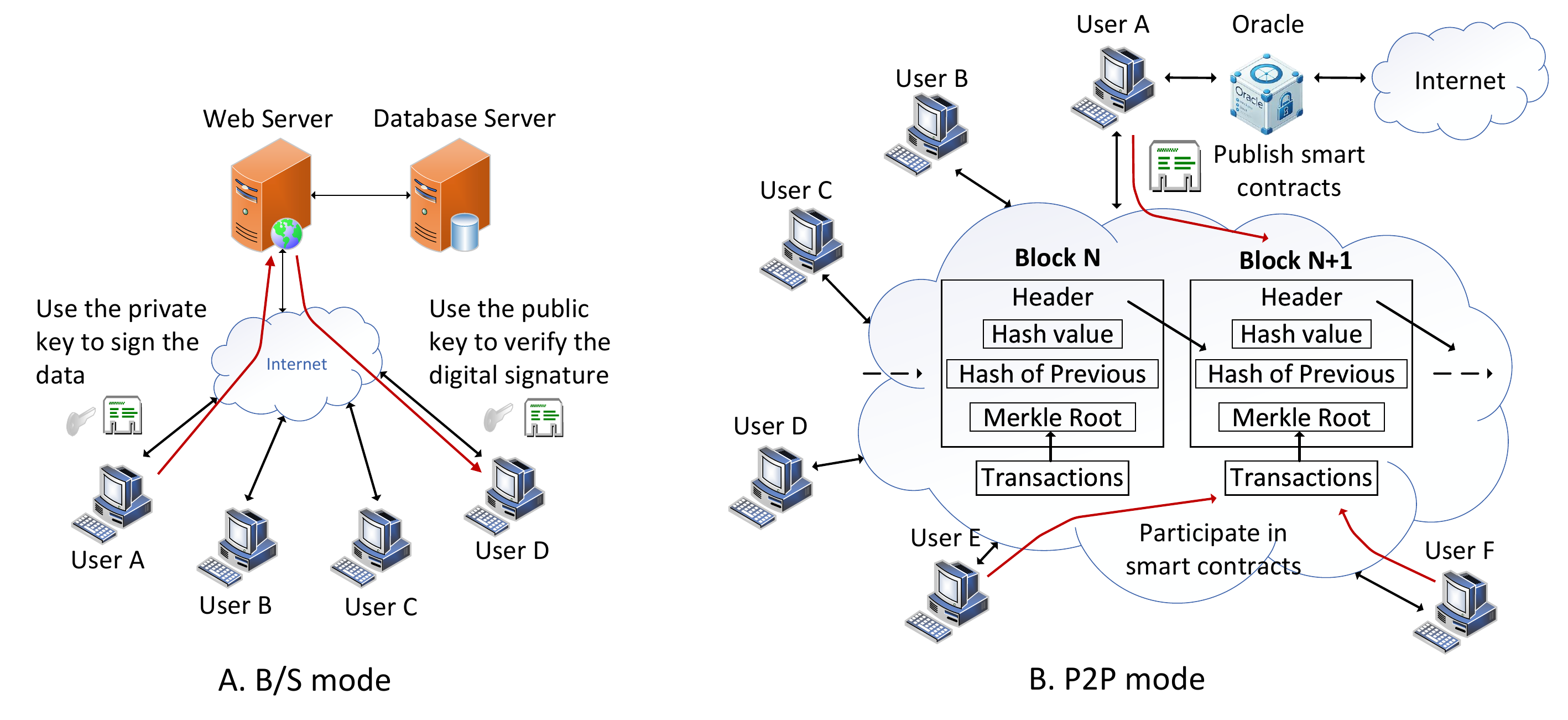}}
	\caption{Structural comparison between B/S and P2P communication modes.}
	\label{fig:ICT carbon market}
\end{figure*}

\subsubsection{Carbon trading market}
There are two main trading mechanisms in the carbon market: cap-and-trade and carbon credit trading.

Cap-and-trade sets a mandatory carbon emission cap for businesses, allowing them to acquire allowances via government distribution or auctions. Carbon credit trading, through carbon offset mechanisms, enables entities to compensate for emissions by investing in reduction projects like renewable energy or carbon sequestration, generating tradable credits. Carbon credits differ from allowances but can be replaced by them under certain rules. For example, in China, the offset ratio of carbon credits (CCER) used for quota settlement cannot exceed 5\% of the required carbon emission allowances \cite{mee2021}. If a company's free allowances are insufficient, it can buy additional allowances or credits through carbon markets to meet its emission targets, as shown in Fig.\ref{fig:carbon_market}.

Carbon allowance and credit prices are established by government policy, but are also influenced by market dynamics. Allowances, limited in number, usually have higher prices than credits, which come from voluntary projects and carry more uncertainty due to variable outputs \cite{narassimhan2017carbon}. Excess credits can lower allowance prices, prompting strict regulations on credit usage and limits on their contribution to emission reductions. The carbon market also includes derivatives such as futures and options, similar to other financial markets.

Carbon sink projects participate in the carbon trading market through carbon credit trading. Carbon credit trading is divided into two markets: the Compliance Carbon Market and the Voluntary Carbon Market. The Compliance Carbon Market, governed by national or international regulations, helps businesses and countries meet mandatory emission reduction targets and includes legally obligated entities. Examples include the Clean Development Mechanism (CDM), which generates Certified Emission Reductions (CER) \cite{IPCC2018}, the EU ETS, and China's Carbon Trading System. In contrast, the Voluntary Carbon Market trades Voluntary Emission Reductions (VER) and is tailored to smaller businesses, NGOs, and individuals, offering simpler entry. The key distinction between these markets lies in the legal obligations versus the voluntary participation that influence the strategic decisions of investors.

The carbon market is a policy tool that reduces societal emissions costs through carbon pricing. This pricing impacts whether entities reduce emissions or buy allowances and influences the strategic and investment decisions of those selling carbon allowances or credits. Carbon prices stabilize in the secondary market, influenced by market forces and government policies. Supplies are controlled by the government, and prices may not yet reflect true market conditions. In this market, allowances and credits are traded, moving toward equilibrium as the market matures.

Under ideal conditions, a stable carbon price reflects the societal average marginal abatement cost. Companies first adopt abatement measures where costs are below the carbon price, but as reduction becomes more costly, they buy allowances instead. Firms with lower abatement costs sell in the carbon market, while those with higher costs buy, optimizing societal abatement costs while meeting government targets.

\subsection{The role of ICT in carbon sink project economics}
\subsubsection{Framework construction and digitalization for financing and trading}
In recent years, ICT has been widely applied in the carbon trading market through two main architectural approaches. Traditional systems employ a Browser/Server (B/S) architecture (Fig.\ref{fig:ICT carbon market}A), where centralized servers manage transactions using established security protocols including encryption (TLS 1.3) and digital certificates to protect high-value carbon assets and sensitive data \cite{SP800-38D}. These systems enable real-time price updates and secure document transfers, but remain vulnerable to single-point failures and require intermediaries for transaction verification.

\begin{figure}
	\centering
	\resizebox{0.32\textwidth}{!}
    {\includegraphics{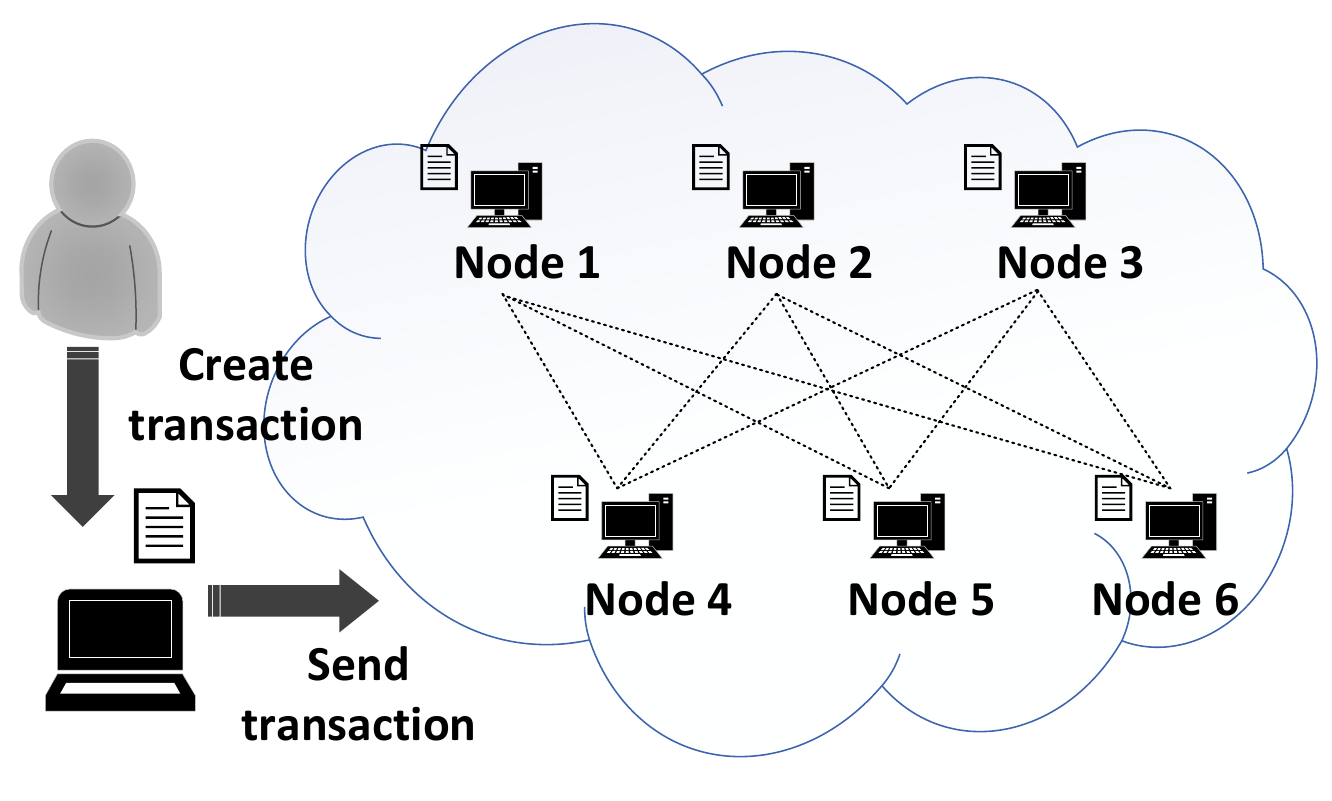}}
	\caption{Structural workflow of smart contracts in a decentralized network}
	\label{fig:Smart}
\end{figure}

The emergence of blockchain technology has introduced a decentralized Peer-to-Peer (P2P) alternative (Fig.\ref{fig:ICT carbon market}B), which has become the main focus of research. Unlike traditional systems, blockchain-based carbon trading eliminates central authorities by distributing transaction records across multiple nodes. Through consensus mechanisms such as PBFT for private networks or DPoS for public chains \cite{mingxiao2017review}, these systems validate transactions without intermediaries. Smart contracts (Fig.\ref{fig:Smart}) automate critical functions, including carbon quota allocation and compliance monitoring by encoding business rules directly into the blockchain \cite{FIPS180-4}. When integrated with real-time data feeds (oracles), these contracts can automatically execute trades based on predefined conditions, significantly improving transparency, reducing transaction costs, and improving market efficiency compared to traditional manual processes.
Although the architectural evolution from B/S (Browser/Server) to blockchain systems has enabled decentralized transaction handling and programmable compliance, real-world carbon trading platforms face broader demands. These include cross-platform interoperability, dynamic regulation enforcement, user-level trust mechanisms, and measurable performance improvements. In response, recent studies have introduced new designs and evaluations aimed at addressing these operational needs and extending the foundational ICT infrastructure of carbon finance.

\paragraph{\bf Theoretical research}
To address the challenges of fragmented registries and inefficient governance, Saraji and Mike (2021) proposed a novel approach: representing carbon credits as Non-Fungible Tokens (NFTs), embedding unique asset identities and programmable trading rules directly into token structures \cite{saraji2021blockchain}. This design enables carbon assets to be traced across platforms without centralized coordination, addressing the demand for interoperability and traceability in decentralized carbon markets. Their key contribution lies in reimagining carbon credits not merely as ledger entries but as self-contained, enforceable digital instruments, laying a foundation for asset-level automation.

Building on this concept of programmable carbon assets, Malamas et al. (2024) expanded the framework into a broader financial infrastructure that encompasses green bonds and other instruments linked to ESG \cite{malamas2024blockchain}. Their model introduced a modular architecture that supports tokenized issuance, embedded compliance validation, and multi-asset compatibility. By extending NFT-based representation to a cross-domain financial system, they contributed a generalizable, policy-aligned infrastructure that enables integrated supervision and secondary market circulation. This work complements Saraji and Mike by broadening the scope of the application, while also introducing governance traceability mechanisms to enhance institutional credibility.

These theoretical contributions collectively reflect a developmental trajectory: from token-level interoperability to cross-asset financial infrastructure. Both works resonate with the architectural emphasis of the current study on smart contract automation and data-driven compliance, providing the conceptual foundation for our performance evaluation–oriented framework.

\paragraph{\bf Empirical research}
The effectiveness of these blockchain-based architectural designs has also been tested and verified in real-world deployments. These empirical studies confirm that theoretical innovations not only improve system design, but also yield measurable gains in trust, speed, and transparency.

Zhang et al. (2024), examining the Shanghai Environment and Energy Exchange, conducted one of the first real-world assessments of blockchain-enabled carbon trading \cite{zhang2024blockchain}. By analyzing the first certified blockchain transaction in April 2022, they found that the transaction speed improved by 40\%, the transaction costs decreased by 15\%, and the abnormal returns reached 8.67\% within 15 trading days. Their study offers robust empirical evidence that decentralized trust and smart contract-based automation can directly boost market performance.

Dong et al. (2024) translated these architectural mechanisms into a system-level simulation, modeling blockchain adoption in China's provincial carbon markets \cite{dong2024explore}. Using an S-shaped diffusion framework that captures firm-level behavioral dynamics under policy and technical incentives, they found that blockchain integration alone can improve overall market effectiveness by up to 20\%. This work highlights the interplay between digital governance and policy structures and its role in shaping adoption outcomes.

Complementing these, Boumaiza and Maher (2024) implemented a localized peer-to-peer carbon trading system in Qatar’s Education City \cite{boumaiza2024leveraging}. Their system integrated real-time emissions data via IoT devices with automated compliance through smart contracts. The resulting platform not only ensured transparency and user trust, but also demonstrated how blockchain-based tools can be adapted to regional regulatory contexts and operational needs.

Together, these empirical studies reinforce the practical value of theoretical blockchain frameworks in real market settings. They demonstrate that innovations such as tokenized carbon assets, embedded compliance logic, and interoperable registries can materially improve the transparency, efficiency, and trustworthiness of carbon trading systems - core goals echoed in the design of the present study.

\subsubsection{Forecasting carbon data trends}

Accurate forecasting of carbon prices plays a pivotal role in enabling informed decisions within the trading and financing markets. While the architectural advancements discussed earlier have laid the groundwork for reliable data collection and infrastructure, new forecasting methods must address complex temporal dependencies, limited sample sizes, and volatile multisource signals inherent to carbon markets. These challenges have stimulated research into advanced deep learning frameworks that offer enhanced robustness and generalization. In what follows, we review recent efforts, structured into theoretical research and empirical validation, that aim to meet these evolving forecasting needs.

\paragraph{\bf Theoretical research} 
To address the non-stationary and nonlinear characteristics of carbon price series, Yang et al. proposed a hybrid model that integrates multifactor ensemble empirical mode decomposition (MEEMD) with LSTM networks \cite{min2022carbon}. This design enables finer-grained feature extraction and temporal pattern learning. Their innovation further lies in optimizing the inference process using a production rule-based reasoning system, improving interpretability and adaptability. Building on the limitations of standard deep models for handling sparse data, Zhang and Wen introduced a novel TCN-Seq2Seq architecture tailored for few-shot carbon market prediction \cite{zhang2022carbon}. Their contribution lies in leveraging fully convolutional temporal layers and a sequence-to-sequence structure to enhance prediction stability with limited data. These two studies share a common trajectory: they both go beyond traditional statistical learning by integrating signal decomposition, deep memory, or sequence modeling, and together reflect an emerging consensus on hybrid learning as the path forward for carbon forecasting in data-sparse, high-noise settings.

\paragraph{\bf Empirical research}
Recent empirical works have demonstrated the real-world effectiveness of these theoretical advances. For example, Huang and Zhang developed a multi-source information framework based on conditional GANs to forecast prices in China’s Hubei and Guangdong pilot carbon markets \cite{huang2024forecasting}. Their framework uniquely integrates macroeconomic, energy, and historical price signals as conditional input and achieves significantly lower MAE and MAPE scores compared to benchmark models. These results highlight the advantage of adversarial learning in capturing dynamic dependencies and enhancing multistep forecast accuracy. Complementing this, Zhang et al. proposed a hybrid model VMD-CNN-BiLSTM-MLP to predict the prices of the EU carbon futures \cite{zhang2023predicting}. The integration of variational mode decomposition enables the model to disentangle high-frequency noise, while the deep-stacked architecture facilitates robust long-horizon forecasting. Their empirical analysis demonstrates consistent improvements in $R^2$ and MAPE across multiple time frames, reinforcing the potential of hybrid deep learning systems in volatile carbon markets.

\subsubsection{ICT-enhanced efficiency and market effectiveness in carbon trading}
The previous discussion illustrated how ICT technologies, particularly blockchain- and AI-based forecasting models, have been applied to improve carbon trading systems. These applications have led to measurable improvements in several key performance indicators (KPIs), which form the basis for an emerging framework to evaluate ICT contributions in this domain. 


Firstly, transaction efficiency, which encompasses transaction speed and cost, has seen significant improvement. For example, the deployment of blockchain at the Shanghai Environment and Energy Exchange resulted in a 40\% increase in transaction speed and a 15\% reduction in transaction costs. These improvements reflect the advantages of automation and peer-to-peer design, which reduce procedural overhead and eliminate intermediaries. Secondly, the effectiveness at the market level, measured by market coordination and overall performance, increased by up to 20\% with blockchain adoption, as shown in simulation studies by Dong et al. The transparency and verifiability provided by blockchain smart contracts played a central role in achieving these gains.

In addition to transaction efficiency and market effectiveness, predictive accuracy has also improved with the integration of AI models for carbon price forecasting. A deep learning hybrid model, which achieved a root mean square error (RMSE) of 0.1960 and a mean absolute percentage error (MAPE) of 2. 61\%, outperformed traditional forecasting methods, demonstrating the potential of AI to provide more reliable decision support.

These improvements in transaction efficiency, market effectiveness, and prediction accuracy serve as primary indicators to evaluate the impact of ICT in carbon finance systems. Future research should aim to standardize these metrics and explore the trade-offs between blockchain and traditional database systems, especially with respect to their energy efficiency and scalability in large-scale applications.

\subsection{Gaps in ICT for carbon sink economics}
\subsubsection{Data compatibility challenges in blockchain-based carbon credit trading}
Blockchain technology has emerged as a promising solution for improving transparency and efficiency in carbon credit trading markets. In particular, the environmental concerns associated with early blockchain consensus mechanisms, such as Proof of Work (PoW), have been substantially mitigated. Contemporary carbon trading platforms predominantly use energy-efficient consensus algorithms such as Proof of Stake (PoS), Delegated Proof of Stake (DPoS), and Zero-Knowledge Proofs (ZKP) \cite{ferdous2020blockchain, zheng2017overview}, which significantly reduce energy consumption and carbon emissions compared to PoW-based systems.

Despite advances in consensus mechanisms, persistent challenges in data compatibility and interoperability hinder seamless interaction between diverse blockchain platforms and traditional carbon credit registries. The lack of standardized protocols limits the cross-platform recognition and tradability of tokenized carbon credits, fragmenting the global carbon market. In addition, integrating blockchain solutions with legacy systems faces technical and operational hurdles due to disparities in data formats, verification processes, and regulatory compliance requirements. These inconsistencies in data exchange and synchronization risk undermining the reliability of carbon trading mechanisms.

\subsubsection{Shortcomings in ICT research for carbon sink financing}

Carbon sink financing is a critical component of green financing, but it has distinct characteristics compared to other forms of green financing.

Firstly, carbon sink financing has a lower entry barrier compared to traditional green financing, which is typically dominated by institutional investors. By using mechanisms like carbon credit subscriptions, carbon sink projects, such as forest carbon sinks, can attract a wider range of investors, including ordinary citizens. This creates a need for more efficient financing platforms capable of managing small investments from a large number of participants, ensuring optimal fund utilization and fair benefit distribution.

Secondly, carbon sink financing differs from traditional green financing in that its returns are closely tied to the carbon credit mechanism. Compared to traditional green financing, which supports projects with clear and direct return models \cite{ozili2022green}, carbon sink financing involves longer payback periods and is influenced by factors such as carbon sequestration, credit generation, and market price fluctuations. This complexity increases risk assessment challenges. Carbon sink financing also faces greater uncertainties, such as market volatility and policy changes, but offers stronger environmental and long-term social benefits. As a result, financing platforms must have robust risk assessment and management capabilities to navigate these complexities.

Related financing methods have attracted growing academic interest, particularly in the integration of artificial intelligence to enhance the credibility and efficiency of green financial instruments. A notable contribution comes from Hemanand et al. \cite{hemanand2022applications}, who developed an intelligent Financial Maximally Filtered Graph (FMFG) algorithm to advance green finance analytics for sustainable development. Their AI-powered framework synergistically combines ESG data processing with natural language understanding capabilities to assess environmental project viability, achieving exceptional performance (98.85\% accuracy) that surpasses conventional neural network approaches. The model's sophisticated multilayer architecture enables comprehensive analysis of financial time-series data while effectively addressing critical decision-making gaps in green finance, particularly for renewable energy and pollution control investments.

Further expanding this research trajectory, Peng et al. \cite{peng2024artificial} conducted a comprehensive empirical investigation of the role of AI in fostering the coordinated development of green finance and the real economy in 30 Chinese provinces. Through the construction of subsystem indices and application of a coupling coordination model, their study leverages AI patent data and industrial robot penetration rates as key indicators of technological innovation and practical implementation. The findings robustly demonstrate AI's dual capacity to directly mitigate information asymmetries, improve credit accessibility, and optimize financial resource allocation, while indirectly promoting systemic coordination through enhanced institutional quality, technological capabilities, human capital development, and government governance effectiveness. Importantly, rigorous robustness checks and heterogeneity analyses reveal these effects to be particularly significant in regions with more advanced AI infrastructure and investment, underscoring AI's transformative potential in creating a more integrated and efficient green financial ecosystem.

\subsubsection{Limitations of carbon data trend prediction models}
From existing research works on carbon data prediction models, it is evident that deep learning neural networks, such as LSTM and Transformer \cite{vaswani2017attention}, have been widely applied to carbon data prediction due to their strong ability to handle sequential data. This is undoubtedly a positive trend, as these models have significant advantages in capturing temporal dependencies and patterns. However, compared with current frontier research in time series prediction models, those applied to carbon data prediction still appear relatively rudimentary.

Currently, state-of-the-art (SOTA) models such as iTransformer \cite{liu2023itransformer}, PatchTST \cite{nie2022time}, and the latest Bi-Mamba+ \cite{liang2024bi} have shown higher prediction performance in time series forecasting tasks. These models not only extract effective information from time series more efficiently than traditional LSTM and Transformer, but also significantly improve the model's generalization performance by reducing learning from noisy data. Moreover, these models generally exhibit better adaptability, capable of handling longer time series and maintaining good prediction accuracy even in the case of scarce data.

At the same time, we should note that despite the superior performance of many SOTA models, their computational resource requirements are relatively high. In the practical application of carbon data prediction, the development of lightweight and efficient models will become an important research direction, especially in scenarios where real-time monitoring and prediction of carbon markets are required. Furthermore, since carbon data prediction plays a crucial role in policy formulation and market operations, improving the interpretability of prediction models will be key. For example, by revealing which variables have the greatest impact on the prediction outcomes, clear decision-making support can be provided to carbon market participants and policymakers. These directions will undoubtedly be key focus areas for future research.
\begin{figure}[H]
	\centering
	\resizebox{0.5\textwidth}{!}{\includegraphics{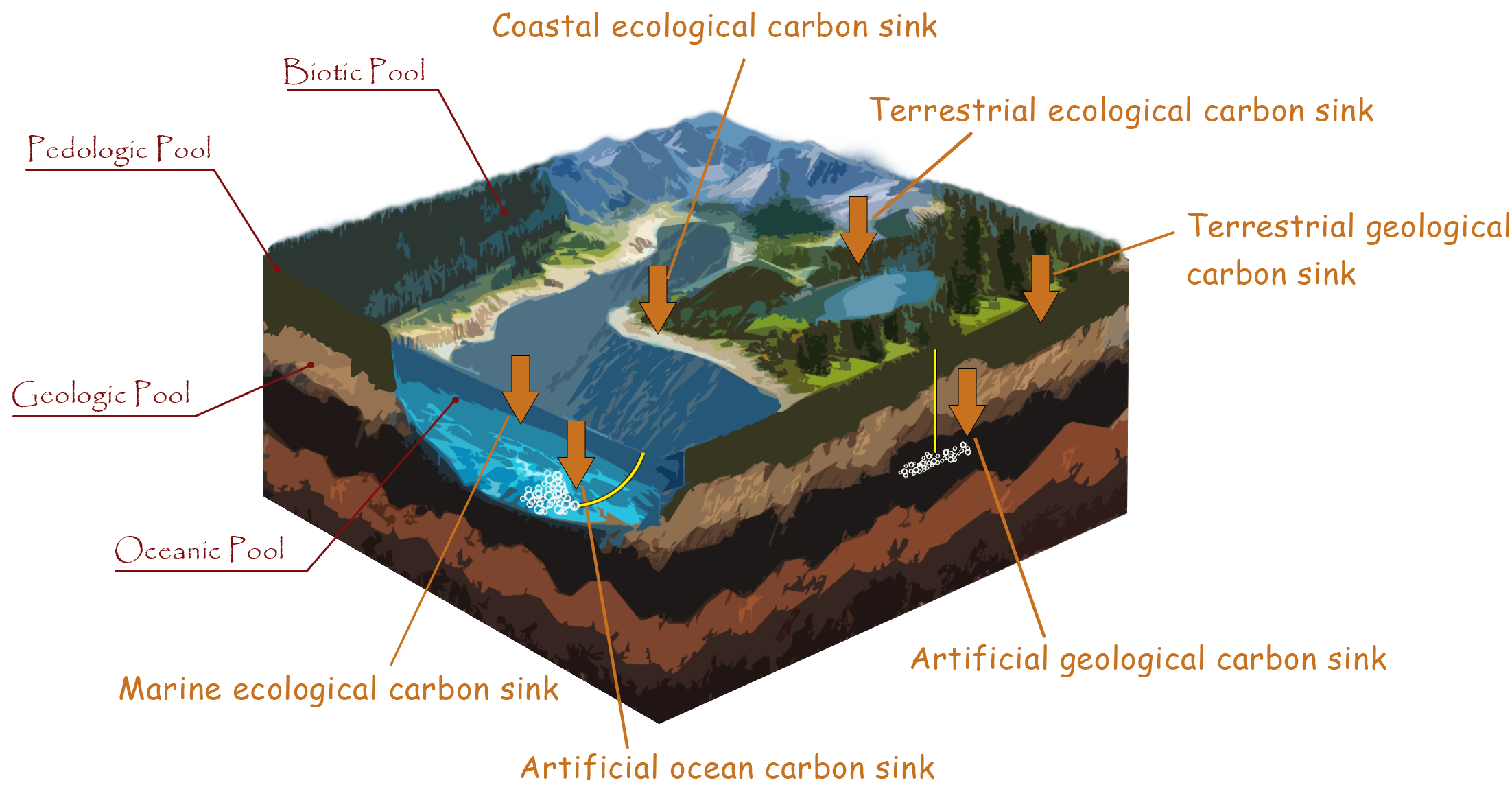}}
	\caption{Schematic diagram of the five major carbon pools and six secondary carbon sink geographic locations. }
	\label{fig:cycyle_and_sink}
\end{figure}

\section{ICT in carbon sink projects: engineering implementation}
\begin{figure*}
	\centering
	\resizebox{0.73\textwidth}{!}{\includegraphics{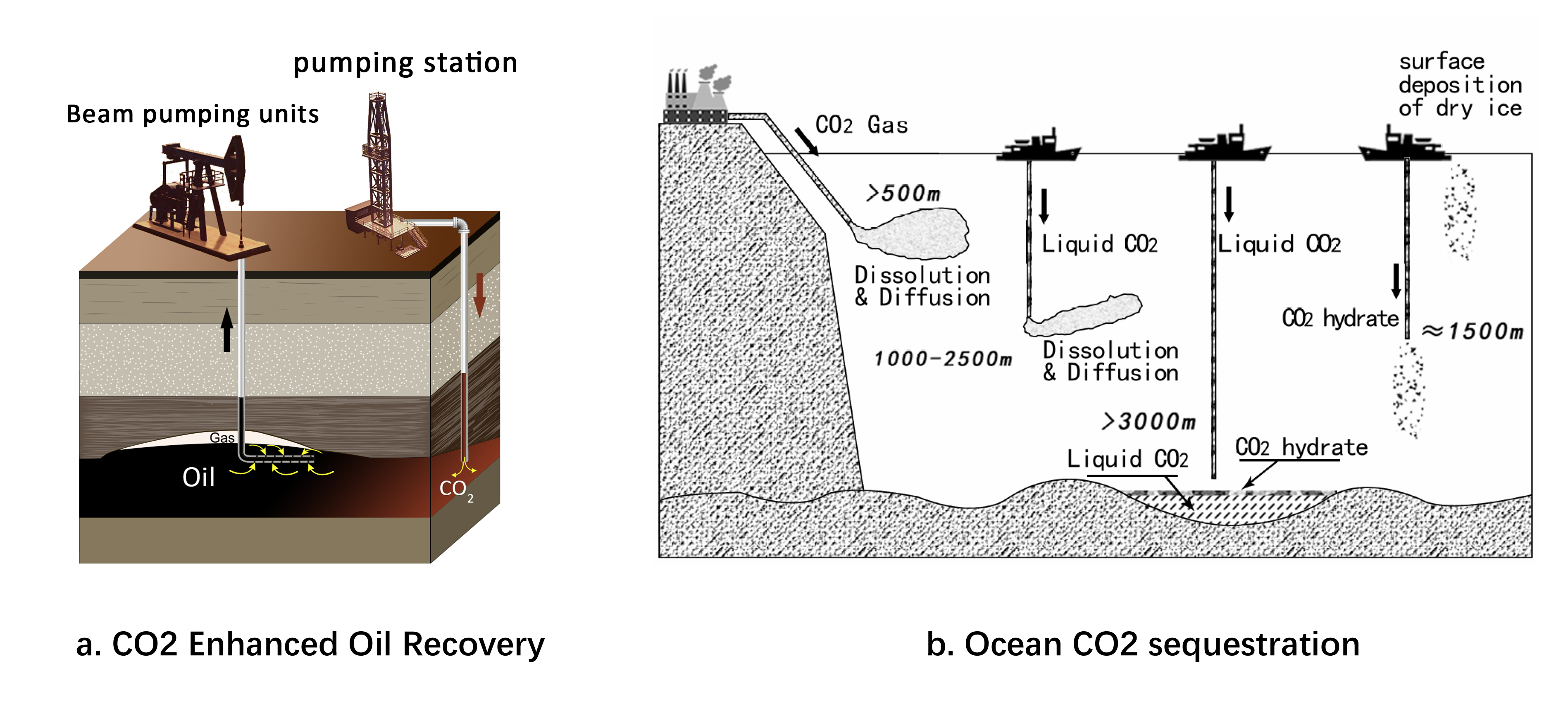}}
	\caption{$CO_2$ storage in geological and oceanic reservoirs.}
	\label{fig:Schematic_diagram_carbon_sequestration}
\end{figure*}
Carbon capture and storage are vital parts of engineered carbon sink strategies. As highlighted in the IPCC Sixth Assessment Report (AR6), large-scale CCS deployment is essential to achieve net zero $CO_2$ emissions, particularly in hard-to-abate sectors \cite{bashmakov2022climate}. Given the technical complexity and long timescales of CCS operations, the report emphasizes the role of digital technologies in improving system integration, monitoring, and operational efficiency. Information and communication technologies are therefore indispensable to ensure the safety, transparency, and long-term reliability of carbon sequestration projects. In this context, the engineering implementation of carbon sink projects  primarily refers to carbon sequestration. For clarity and consistency, the term "carbon sequestration" will be used throughout this section.

\subsection{Basic implementation methods of carbon sequestration}
There are five primary carbon reservoirs worldwide: the oceans, atmosphere, soil, geological formations, and biological carbon pools, as illustrated in Fig.\ref{fig:cycyle_and_sink}. The carbon exchanges among these reservoirs collectively constitute the global carbon cycle. From a geographical perspective, carbon sinks can broadly be classified into terrestrial and oceanic carbon sinks.

Terrestrial carbon sinks involve mechanisms and processes that absorb atmospheric $CO_2$ through terrestrial ecosystems and geological activities, which encompasses biological, geological and soil carbon pools. These terrestrial carbon sinks can be subdivided into terrestrial ecological carbon sinks, terrestrial geological carbon sinks, and artificial geological carbon sinks \cite{weidong2023advances}. 

Oceanic carbon sinks absorb atmospheric $CO_2$ through marine processes and biological activities, involving both oceanic and biological carbon pools. These oceanic sinks can be classified into coastal ecological carbon sinks, oceanic ecological carbon sinks, and man-made oceanic carbon sinks \cite{weidong2023advances}.

The essence of the carbon sink project is to promote the absorption of carbon from the atmospheric carbon pool by other carbon pools through engineering means, thus reducing atmospheric concentrations $CO_2$. Currently, typical implementation methods can be categorized into nature-based carbon sequestration and engineering-based CCUS.

\subsubsection{Enhanced natural carbon sequestration}

Enhanced natural carbon capture involves the use of technology to increase carbon storage in ecosystems, helping to lower atmospheric carbon dioxide. This overview highlights two main types: forest carbon sinks for land and phytoplankton cultivation for oceans. 

Forests play a vital role as carbon sinks, sequestering about $2.4 \pm 0.4 , Pg , C , year^{-1}$ from 1990 to 2007. Key methods include afforestation (AF), reforestation (RF), and improved forest management (IFM). AF involves planting trees in areas that have not been used for more than 50 years, while RF restores forests in recently deforested areas \cite{pearce2019green}. Both increase carbon storage, but can disrupt ecosystems if poorly implemented \cite{portner2022ipcc}. IFM improves forest health and carbon sequestration through better management practices \cite{Moomaw2019IntactFI, putz2008improved} . While forest carbon sinks help mitigate climate change, they cannot fully restore preindustrial carbon levels. \cite{Moomaw2019IntactFI}.

The carbon sinks of ocean aquaculture involve the cultivation of seaweed and seagrass to absorb atmospheric $CO_2$, the carbon being sequestered in the deep ocean for thousands of years \cite{Ortega2019ImportantCO}. Marine phytoplankton, while making up only 1\% of global plant biomass, contributes to half of global carbon fixation, highlighting its potential for large-scale ocean afforestation \cite{behrenfeld2014climate}. Unlike terrestrial carbon sinks, such as forests, ocean-based carbon sinks do not compete for land resources, making them easier to implement. It is important to note that phytoplankton cultivation for commercial products such as food and biofuels offers only temporary carbon absorption, while carbon sink-focused cultivation aims for long-term sequestration by absorbing $CO_2$ and depositing it in the deep sea, often depending on carbon credits and government subsidies for profitability.

\subsubsection{Carbon capture, utilization, and storage}
CCUS, compared to approaches that improve natural carbon capture, directly targets high concentration $CO_2$ emission sources for efficient and rapid capture and treatment, significantly reducing carbon emissions from human activities.

$CO_2$ capture involves the separation of carbon dioxide from power plants, industrial facilities or the atmosphere, using methods such as post-combustion and pre-combustion \cite{nagireddi2023carbon}. Key separation techniques include adsorption, membrane separation, and cryogenic technology \cite{das2023review}. For transport, pipelines are used for large-scale transfers, while ships and tankers handle offshore and short-route logistics.
\begin{figure*}
	\centering
	\resizebox{0.85\textwidth}{!}{\includegraphics{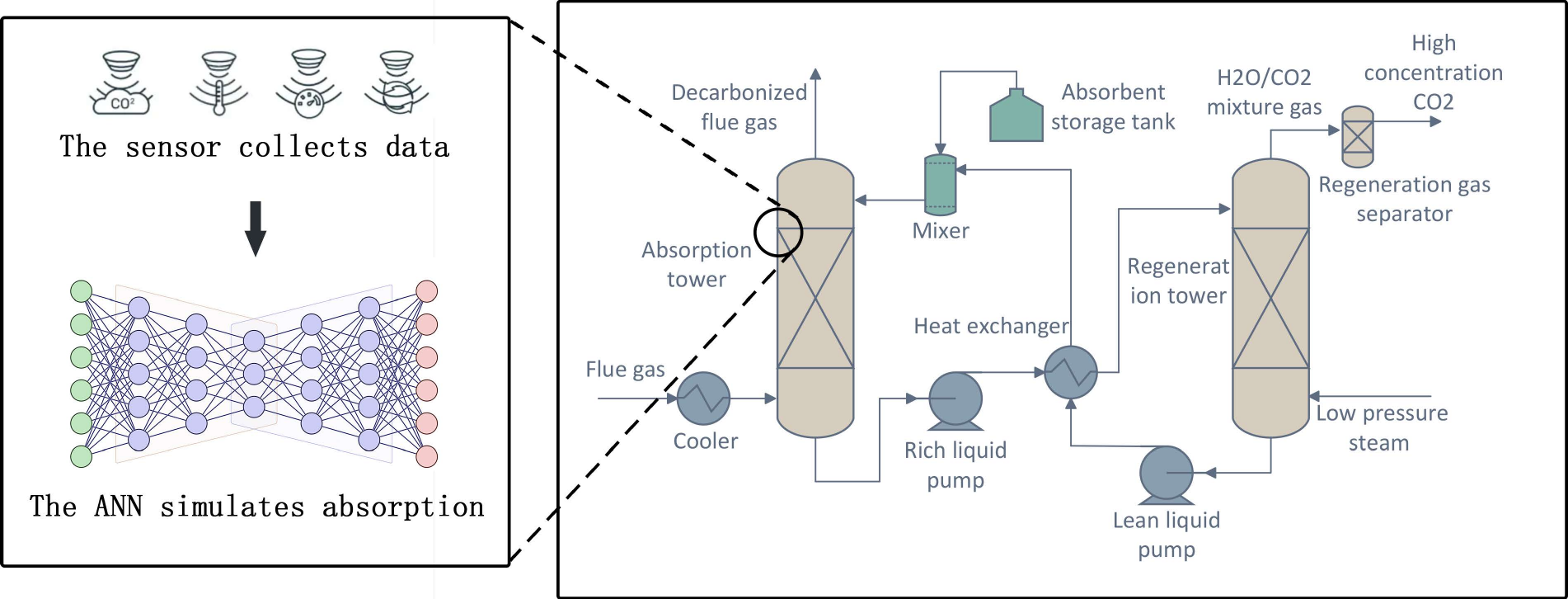}}
	\caption{Application of artificial intelligence in carbon capture.}
	\label{fig:absorption and regeneration}
\end{figure*}
$CO_2$ is the main culprit of global warming, but it also has great industrial value. With the development of technology, its uses are divided into physical and chemical applications. Physically, the use of $CO_2$ is based on its phase change properties, converting it into a useful form or using its properties as a medium without changing its chemical structure. The uses of gaseous $CO_2$ include oil recovery, promoting the growth of greenhouse plants \cite{prior2011review}, and even starch synthesis \cite{cai2021cell}; liquid $CO_2$ can be used as an environmentally friendly cleaner and refrigerant \cite{belman2014general}, replacing harmful chemicals; solid $CO_2$ or dry ice can be used in cold chain logistics in the transport of food and drugs and artificial rainfall; supercritical $CO_2$ is used in extraction processes \cite{sahena2009application}, cleaning, and high pressure washing. Chemically, the conversion of $CO_2$ into useful products involves chemical reactions, such as the production of chemical fertilizers, especially urea. Although using $CO_2$ to create products is promising, limitations in demand, technology and regulations mean direct storage is still essential, with enhanced oil recovery and ocean storage being common storage methods.

$CO_2$ Enhanced Oil Recovery (EOR) entails the injection of $CO_2$ into exhausted oil reservoirs to increase oil extraction, applicable to both terrestrial and offshore fields. Initially, the pressure of the natural reservoir drives extraction, followed by the injection of water to maintain the pressure when it drops. However, because of differences in permeability and viscosity between oil and water, water often bypasses oil, leading to poor recovery and the phenomenon of water breakthrough. The typical recovery from these first two phases ranges from 20\% to 40\% \cite{EPRI1999}. When water flooding becomes economically inefficient, tertiary recovery \cite{alvarado2010enhanced} begins. In $CO_2$ EOR, supercritical $CO_2$ dense as a liquid and viscous as a gas penetrates the oil layer, mixing with and thinning the oil, thus improving its flow. It also dissolves and expands the oil, driving it toward production wells \cite{nagabhushan2016emission}. This method can theoretically recover all the oil contacted, but practically yields an additional recovery 10\%-15\% \cite{EPRI1999}. $CO_2$ EOR is particularly effective in high viscosity or low permeability reservoirs, and most of the injected $CO_2$ remain sequestered for the long term, offering economic and environmental advantages. A graphical representation is shown in Fig.\ref{fig:Schematic_diagram_carbon_sequestration}(a).

Sequestration of the ocean $CO_2$ involves injecting $CO_2$ into the sea at various depths to isolate it from the atmosphere, as shown in Fig.\ref{fig:Schematic_diagram_carbon_sequestration}(b). There are three main methods. Firstly, gas injection involves the release of $CO_2$ gas at depths greater than 500 meters, where it rapidly dissolves and disperses in seawater \cite{caldeira2005ocean}. Secondly, liquid injection uses liquid $CO_2$, which is injected below 500 meters, typically deeper than 1000 meters \cite{saito2000co2}. Between 500 and 2500 meters, where the density of liquid $CO_2$ is less than that of seawater, it forms a rising plume that gradually dissolves \cite{caldeira2005ocean}. Below 3000 meters, its density exceeds that of seawater, causing it to sink and potentially accumulate in seafloor depressions, thus forming 'carbon lakes' \cite{caldeira2005ocean}. Lastly, solid injection and hydrates involve dispersing dry ice on the surface, which then sinks to the seabed and dissolves slowly. Alternatively, $CO_2$ hydrates, which are stable solids formed under high pressure and low temperatures, can be injected at specific depths, around 1500 meters, using pipelines \cite{CHOW20094937}.

\subsection{The roles of ICT in carbon sequestration implementation}
Current research focuses primarily on modeling the various stages of carbon sequestration, particularly through the use of deep learning models, as well as leveraging digital technologies for the efficient processing of large datasets.

\subsubsection{Deep learning in carbon sequestration}

Traditional carbon sequestration modeling and control methods, such as empirical calibration and physics-based simulations, often struggle to manage the nonlinear, multivariable, and dynamic characteristics of real-world systems. These limitations, coupled with the growing availability of operational and experimental data, have spurred increasing interest in deep learning as a more flexible and adaptive approach. Recent research in this area can be broadly categorized into two directions: one focusing on enhancing carbon capture, and the other on supporting carbon storage.

For carbon capture, efficient absorption of $CO_2$ is critical. As illustrated in Fig.\ref{fig:absorption and regeneration}, current technologies rely on physical or chemical absorption, where $CO_2$ is separated from flue gases using solvents such as polyethylene glycol dimethyl ether in physical methods, or alkaline solutions in chemical methods \cite{yu2012review}. However, optimizing these processes remains difficult because of the non-linear dynamics of reaction kinetics, fluid flow, and energy coupling, making them ideal candidates for data-driven modeling.

In contrast, carbon storage focuses on monitoring $CO_2$ plume migration and assessing leakage risk within deep geological formations. These tasks involve interpreting complex subsurface dynamics, which are typically governed by multiscale interactions among injection pressure, reservoir heterogeneity, and capillary effects. Accurate prediction of these phenomena poses substantial challenges for traditional simulation tools, particularly under uncertainty or sparse sensor coverage.

\paragraph{\bf Theoretical research}
Recent theoretical work has focused on designing deep learning models that can simulate complex carbon sequestration processes while preserving physical accuracy. Du et al.~ \cite{du2023modeling} proposed a physics-informed neural network (PINN) that combines domain decomposition with Fourier features. This design improves the accuracy of the simulation for CO$_2$ movement in porous media while greatly reducing the computational cost, offering a practical alternative to traditional numerical solvers.

Kadeethum et al.~ \cite{kadeethum2023efficient} introduced a domain-decomposed neural operator (DDNO) that learns flow dynamics across different spatial regions. Unlike standard neural networks, DDNO captures the full solution process and includes a built-in uncertainty estimation method, helping improve model reliability and generalization. These models advance the theoretical foundations of deep learning for carbon sequestration by combining physical laws with efficient learning mechanisms.

\paragraph{\bf Empirical research}
Across the carbon sequestration chain, deep learning has shown tangible improvements. In capture, Mohan et al.~ \cite{mohan2023accurate} used ANN with COSMO-RS descriptors to predict CO$_2$ solubility in deep eutectic solvents, reducing the prediction error from 23. 4\% to 2. 7\%, improving solvent screening. For transport, the Petra Nova project~ \cite{DOE_Petra_Nova_2017,dhruv2024ai} applied RL to optimize routing and scheduling, cutting energy and logistics costs by 15\%. 

In storage, Wen et al.~ \cite{wen2021towards} introduced RU-Net for the fast prediction of CO$_2$ plume dynamics in heterogeneous formations, achieving strong generalization through transfer learning. Sinha et al.~ \cite{sinha2020normal} combined ConvLSTM and pressure data to automate leakage detection, addressing monitoring latency. Wang et al.~ \cite{liu2024spatio} further advanced this by using spatiotemporal GNNs to track CO$_2$ migration from sparse satellite imagery, enabling remote and scalable monitoring. 

Complementing these, Dhruv et al.~ \cite{dhruv2024ai} provided a quantitative survey of AI benefits in the stages of CCS, and two domestic studies~ \cite{tavakolian2024modeling,ali2024prediction} validated AI effectiveness in the monitoring of China's saline aquifer. These empirical results collectively demonstrate how AI improves accuracy, efficiency, and responsiveness in carbon sequestration operations.

\subsubsection{Digital technologies driving carbon sequestration}
Beyond model-driven learning paradigms, a parallel line of research focuses on digital infrastructures—such as IoT systems, simulation platforms, and cyber-physical integration frameworks—that aim to enhance the full carbon sequestration lifecycle. These technologies are particularly suited to address challenges in large-scale system coordination, heterogeneous data integration, and real-time operational control, which are often beyond the scope of purely algorithmic methods.

\paragraph{Theoretical research}
To address the lack of unified digital architecture for multi-scale carbon sequestration processes, Fang et al.~ \cite{fang2025digital} developed a conceptual framework synthesizing seven categories of digital technologies—including remote sensing, GIS, and digital twins—mapped to functions such as carbon sink estimation and emission balance analysis. Their theoretical contribution lies in bridging ecological modeling with spatial data infrastructure, providing a foundation for systematic UGS-based carbon assessment.

Focusing on industrial-scale capture systems, He et al.~ \cite{bielka2023future} proposed a digital-twin-based control architecture integrating physical simulation, sensor feedback, and control theory. This work addresses the limitation of reactive control in traditional CCS systems by enabling predictive regulation and multi-layer decision-making, thus offering a scalable template for cyber-physical system design in carbon infrastructure.

\paragraph{\bf Empirical research}
On the empirical front, several implementations have validated the advantages of digital platforms in practical carbon sequestration settings. To tackle inefficiencies in manual monitoring in subsurface injection, Kang et al.~ \cite{chawla2023iot} deployed an IoT-based sensing network that achieved real-time monitoring of CO$_2$ plume dynamics and automated leakage alerts, significantly improving safety responsiveness. Their innovation lies in the integration of low-power sensors with edge computing, enabling continuous and distributed monitoring in complex underground environments.

Zhao et al.~ \cite{valli2024sustainable} applied a similar IoT-enhanced architecture in oil recovery scenarios, improving operational transparency and system uptime. The key contribution is the incorporation of cloud communication and anomaly tracking into the injection workflow, forming a closed-loop architecture that enhances the resilience of the system and the efficiency of carbon utilization.

To address the bottleneck of fragmented data workflows in CCUS operations, Wang et al.~ \cite{kozman2024end} built an end-to-end cloud platform for managing heterogeneous CCUS datasets, including geological models and monitoring records. Their innovation lies in unifying data acquisition, modeling, and collaborative access into a scalable infrastructure, supporting large-scale and continuous decision-making in carbon management. Together, these studies demonstrate how digital technologies, different from pure machine learning models, form the backbone of scalable, adaptive, and real-time carbon sequestration systems.

\subsubsection{ICT-driven improvements in efficiency and accuracy in carbon sequestration}
ICT applications in carbon sequestration systems, particularly AI and digital technologies, have brought about significant improvements in key performance indicators (KPIs), such as predictive accuracy, operational efficiency, and safety. These improvements underscore the transformative potential of ICT in optimizing carbon capture and storage operations.

AI-driven models have significantly improved the accuracy of $CO_2$ adsorption predictions. For example, a multilayer perceptron model achieved an $R^2$ of 0.9978, a substantial improvement over traditional methods. This enhanced prediction accuracy enables better process optimization, reduces uncertainty in carbon capture, and improves overall system efficiency. In terms of operational efficiency, AI optimizations in projects such as Petra Nova have reduced energy consumption and transportation costs by 15\%, leading to significant savings. The integration of digital twin technology further improves pipeline safety and operational efficiency by enabling real-time monitoring and early detection of potential risks. This reduced maintenance costs and increased the reliability of the system.

Safety, a crucial aspect of carbon sequestration, has also benefited from digital technologies. AI and sensor networks are now used to monitor the integrity of storage sites and detect potential leakage, contributing to safer and more reliable carbon sequestration operations. These advances not only improve the economic sustainability of sequestration projects, but also enhance their environmental reliability. These improvements in predictive accuracy, operational efficiency, and safety form a foundational framework for evaluating the role of ICT in carbon sequestration. However, more work is needed to standardize these performance metrics and explore the trade-offs between different ICT solutions, such as AI versus traditional modeling techniques, particularly in terms of computational cost, accuracy, and real-time monitoring capabilities.

\subsection{Gaps in ICT for carbon sequestration}
\begin{figure}
	\centering
	\resizebox{0.45\textwidth}{!}{\includegraphics{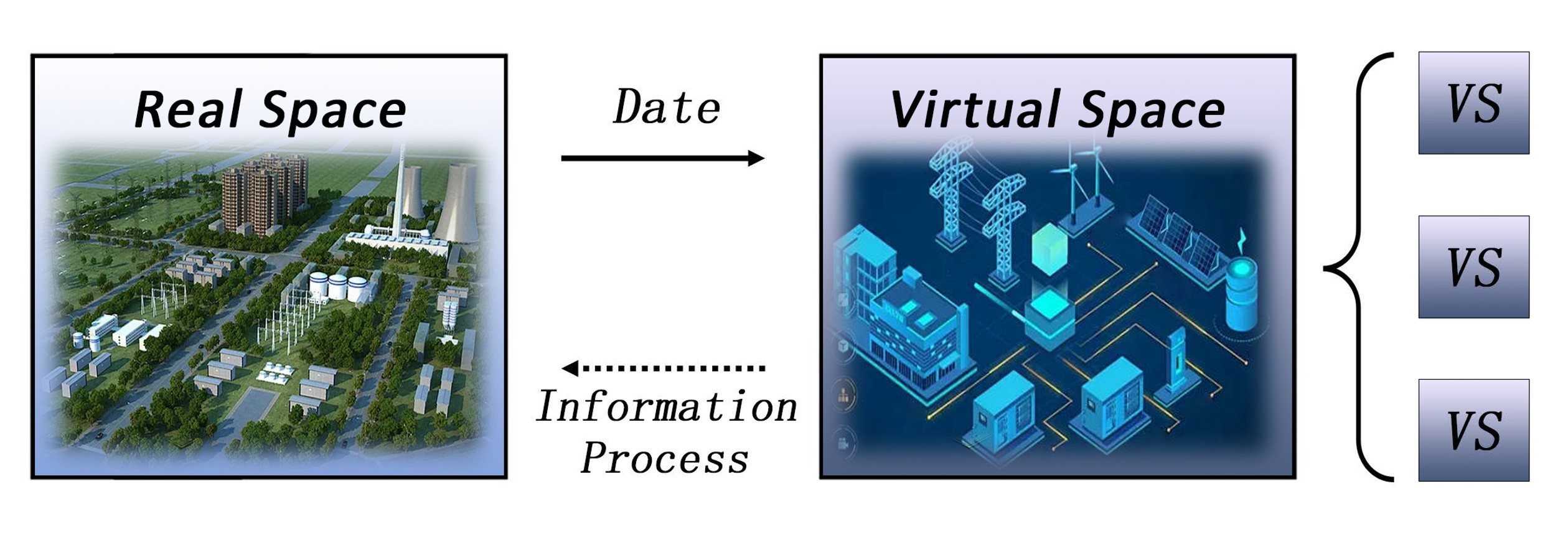}}
	\caption{Digital twin framework for real-time interaction between real and virtual entities.}
	\label{fig:Digital}
\end{figure}
In carbon sequestration, process control is the key to ensuring smooth execution of projects \cite{ravichandran2024carbon}. However, current research primarily focuses on modeling and data analysis. Models for process control, especially those capable of real-time adjustment and optimization of $CO_2$ injection and storage processes, are still underexplored. The actual carbon sequestration process involves the interaction of multiple variables, including temperature, pressure, and permeability, which may change over time and space. How to monitor and adjust the carbon sequestration process in real time under different engineering conditions to ensure its stability and efficiency is an issue that needs to be addressed. Future research should focus more on how to combine control theory with deep learning and optimization algorithms to build real-time feedback mechanisms and optimize the carbon sequestration process.
\begin{figure*}
	\centering
	\resizebox{0.75\textwidth}{!}{\includegraphics{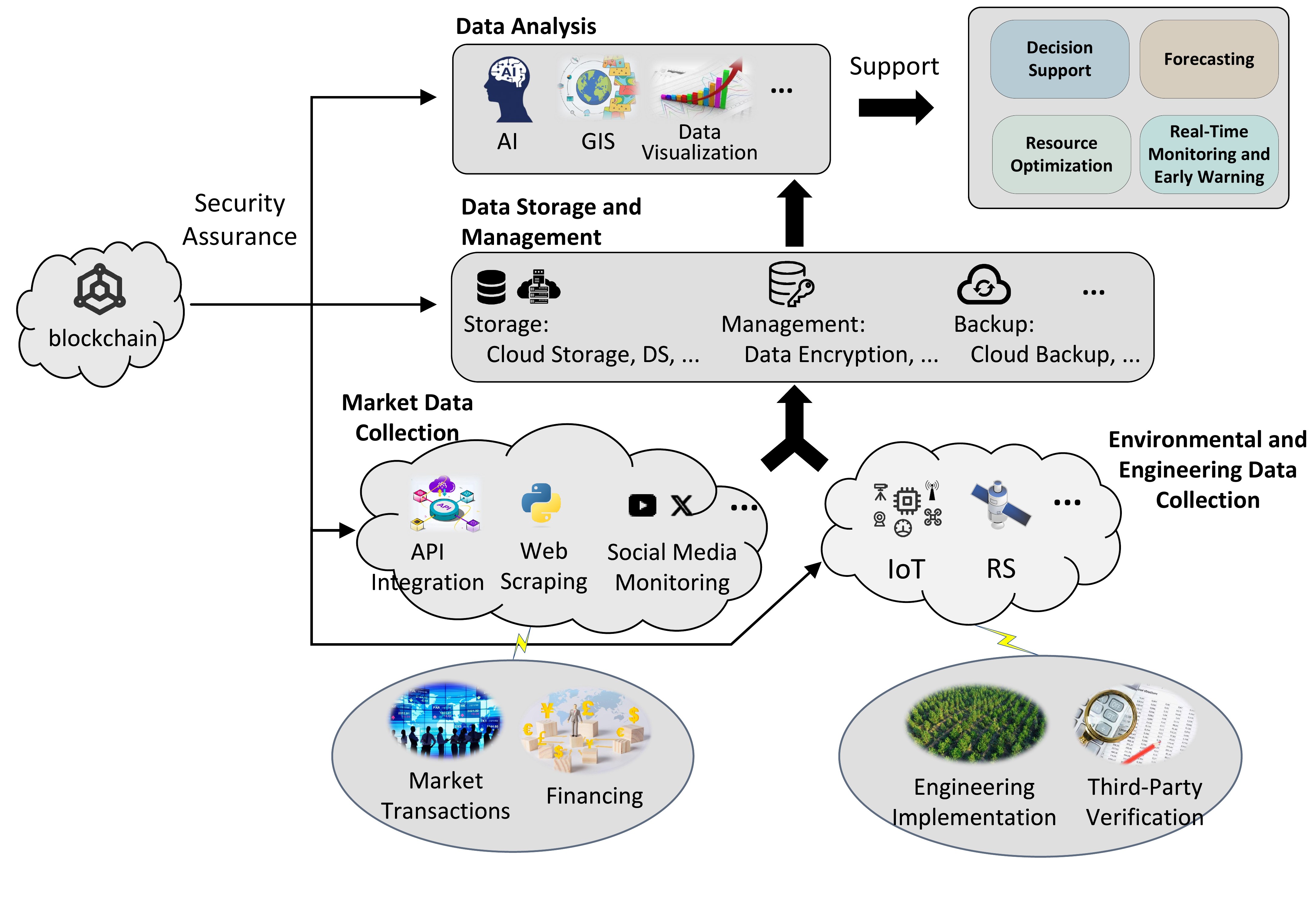}}
	\caption{
		 The ICTs based architecture to promote the carbon sequestration.
		}
	\label{fig:ICT in carbon sink}
\end{figure*}
Digital modeling, especially digital twin(structure illustrated in Fig.\ref{fig:Digital}) and virtual reality technologies, is still in its early stages in carbon sequestration applications. Current digital modeling focuses mainly on specific subsystems, such as $CO_2$ transport and leakage monitoring, but modeling for other key aspects of carbon sequestration, particularly the long-term evolution of underground reservoirs and the interaction between $CO_2$ and underground rock layers, has not been comprehensively developed. Existing technologies need to be further expanded to consider more environmental variables (such as temperature, humidity, and geological changes) and their coupling effects with environmental factors to create more accurate digital models, enhancing their applicability and precision in real-world engineering.

With the increasing volume of data involved in carbon sequestration projects, issues related to data sharing and privacy protection are gradually emerging. Data sharing between different departments and research institutions is essential for advancing carbon sequestration technologies, but ensuring effective data sharing while maintaining data security and privacy remains a significant concern.

Furthermore, while current research has explored the application of various digital technologies in carbon sequestration, such as deep learning and digital twins, these technologies are often applied in isolation to specific problems. In practical applications, how to integrate these technologies effectively so that they work together—such as how to combine digital twins with real-time process control systems or how to optimize reservoir management and $CO_2$ injection strategies through deep learning models—remains a challenge in technology integration. Future research could explore multi-technology integration solutions to improve overall system performance.

At the same time, the application of digital technologies in the promotion and public engagement of CCUS faces several notable challenges. These include the need for substantial time and financial investment, the potential loss of content control in open digital environments, the risks of information overload, the limited interactivity in many current platforms, and the absence of systematic evaluation and feedback mechanisms. Moreover, the coordination of contents among multiple stakeholders can become complex, and reluctance among scientific experts to actively participate in digital communication may further constrain its impact. Addressing these challenges is essential to fully realize the potential of digital technologies in supporting CCUS awareness, education, and policy engagement. These issues need to be addressed by future researchers who should propose more reasonable solutions to promote further improvement in carbon sequestration.
\begin{table*}[htbp]
\centering
\caption{Comparative analysis of representative ICT-enabled carbon sink studies. covering a temporal span from 2021 to 2024 and a wide range of application scenarios including carbon trading, carbon price forecasting, and CO$_2$ sequestration monitoring}
\small
\scalebox{0.95}{
\setlength{\tabcolsep}{4pt} 
\renewcommand{\arraystretch}{1.2} 
\resizebox{\textwidth}{!}{%
\begin{tabular}{
>{\centering\arraybackslash}p{0.4cm} 
>{\centering\arraybackslash}p{4cm} 
>{\centering\arraybackslash}p{4cm} 
>{\centering\arraybackslash}p{4cm} 
>{\centering\arraybackslash}p{5cm} 
>{\centering\arraybackslash}p{5cm}}
\hline
\textbf{No.} & \textbf{Reference} & \textbf{ICT Type} & \textbf{Application} & \textbf{Highlights} & \textbf{Limitations} \\
\hline
1 & Saraji \& Borowczak(2021)~ \cite{saraji2021blockchain} & DLT (smart contracts + NFTs) & Carbon credit tokenization & Decentralized validator-based verification & Testnet prototype only, conceptual design \\
2 & Malamas et al. (2024)~ \cite{malamas2024blockchain} & DLT (ERC-20 + IPFS) & Green bond issuance & Transparent, auditable, reduces intermediaries & Proof-of-concept, high energy consumption \\
3 & Dong et al. (2024)~ \cite{dong2024explore} & Agent-based modeling & China carbon market policy & Simulates blockchain adoption in carbon trading & Simulation only, lacks real-world validation \\
4 & Yang et al. (2022)~ \cite{min2022carbon} & MEEMD-LSTM hybrid & China carbon price forecasting & High $R^2 > 0.98$, robust across markets & Complex preprocessing, requires optimization \\
5 & Zhang \& Wen (2022)~ \cite{zhang2022carbon} & TCN-Seq2Seq & Carbon price forecasting & High accuracy (DA=0.9697, MAPE=0.0027) & Black-box model, limited interpretability \\
6 & Zhang et al. (2023)~ \cite{zhang2023predicting} & VMD-CNN-BILSTM-MLP hybrid & EUA carbon futures forecasting & Optimized decomposition, $R^2 > 0.94$ & Multi-stage complexity \\
7 & Du et al. (2023)~ \cite{du2023modeling} & PINN + Fourier features & CO$_2$ sequestration & Stable long-term prediction, lower cost & Low accuracy in unstable convection \\
8 & Liu et al. (2024)~ \cite{liu2024spatio} & 3D CNN + SimVP & CO$_2$ plume monitoring & No geology input, integrates global-local data & Short-term only, high computational cost \\
9 & Ali et al. (2024)~ \cite{ali2024prediction} & ANN + LSTM & CO$_2$ solubility in ionic liquids & ANN: R²=0.986, faster; large dataset (10k+ points) & High ANN error (28\%), LSTM costly, limited scope \\
10 & Chawla et al. (2023)~ \cite{chawla2023iot} & IoT + fault diagnosis & CCS real-time monitoring & Real-time anomaly detection, sensor validation & Sensor faults, complex diagnostics \\
11 & Valli et al. (2024)~ \cite{valli2024sustainable} & IoT + predictive analytics & CCS-EOR & Optimizes CO$_2$ injection, reduces emissions & High cost, complex infrastructure \\
12 & Kozman et al. (2024)~ \cite{kozman2024end} & FAIR-compliant workflows & CCUS data management & Scalable to petabyte, automated processing & High deployment cost, complex setup \\
\hline
\end{tabular}%
}
}
\label{tab:ict_comparative}
\end{table*}

\section{Conclusion and future directions}
This paper has analyzed that ICT could play vital roles and have great potential throughout carbon sink projects. As shown in Fig.~\ref{fig:ICT in carbon sink}, they support each stage from initial planning to field implementation. In the early phase, ICT enable multi-source data collection from economic systems and engineering sites, followed by structured storage and scalable management. Advanced tools, including AI-based forecasting models and digital decision support systems, help optimize design strategies, improve implementation efficiency, and enable real-time monitoring. Technologies such as blockchain further enhance trust and transparency by ensuring data immutability and secure carbon asset management.

To illustrate the breadth and depth of current ICT-enabled applications in the carbon sink domain, we have compiled a comparative analysis of 12 representative technical studies (Table~\ref{tab:ict_comparative}). These works span blockchain-based carbon market infrastructures, AI-powered carbon price forecasting models, IoT-based carbon storage monitoring systems, and data-driven CCUS lifecycle platforms. The table has synthesized their core ICT types, application domains, methodological innovations, performance advantages, and implementation limitations. This structured comparison not only reflects the diversity and maturity of existing efforts, but also highlights the integrative perspective of our work in bridging financial, engineering, and computational innovations. Although this paper has comparatively analyzed key limitations in current ICT-enabled carbon sink applications, it is important to note that these are primarily technical in nature, such as algorithmic performance, system scalability, and implementation maturity. However, the applications of ICT in this domain would also face broader systemic challenges that go beyond individual technologies and require further research and improvement.

\paragraph{\bf Impact of ICT on the environment}

On one hand, ICT can provide technical support for the implementation of carbon sink projects; on the other hand, its operation itself is a significant source of carbon emissions \cite{belkhir2018assessing}, especially when training data-intensive deep learning models. Training large AI models, particularly tasks that involve massive datasets, often requires substantial energy consumption, resulting in significant carbon dioxide emissions. According to relevant studies, energy consumption during the deep learning model training process even exceeds the carbon footprint of some traditional industrial activities. Therefore, to achieve sustainable development goals, some scholars have proposed the concept of quantifying the carbon footprint of AI, which involves accurately calculating the energy consumption and carbon emissions of AI models to assess their environmental impact \cite{cowls2023ai} \cite{dhar2020carbon}.

T Parcollet and M Ravanelli \cite{parcollet2021energy} further practiced the quantification of the AI carbon footprint in their study, investigating the carbon cost of end-to-end automatic speech recognition (ASR) systems. By calculating the energy consumption and cooling expenses during the training process and converting them into $CO_2$ emissions, they quantified the carbon footprint of the model training. Their experiments found that a state-of-the-art transformer (SOTA) model released 50\% of its total $CO_2$ emissions during training solely to reduce the word error rate (WER) by 0.3. In addition to AI, other ICT applications also accompany significant carbon emissions, such as information retrieval \cite{scells2022reduce} and IoT \cite{benhamaid2022recent}.

However, simply quantifying the carbon footprint of ICT cannot solve the problem of its high carbon emissions. Future research should focus on reducing the energy consumption of models. Among emerging solutions, liquid neural networks (LNNs) show promise due to their biologically inspired, continuous-time dynamics and ability to support incremental learning without full retraining \cite{hasani2021liquid, hasani2018liquid}. In wireless communication tasks, LNNs have demonstrated strong energy-efficiency potential: Wang et al. \cite{wang2024robust} improved spectral efficiency by 4.15\% using only 1.61\% of the computation time, while Zhu et al. \cite{zhu2024robust} achieved up to 46.9\% higher normalized spectral efficiency in beam tracking. Despite such advantages, LNNs remain underexplored in carbon sink applications. Meanwhile, established methods such as pruning and quantization also offer effective means of reducing computational costs. Integrating these techniques with lightweight architectures, such as LNNs, may open new directions for low-power intelligent ICT systems in sustainable carbon sink management.

\paragraph{\bf Policy support}
The government plays a key role in promoting the integration of ICT and carbon sink projects \cite{huang2024impact, haseeb2024enhancing}. Currently, many national and regional policies have not adequately addressed how ICT technologies can facilitate the effective implementation of carbon sink. Therefore, there is an urgent need for greater policy support in this area. Governments can encourage in-depth cooperation between technology companies and carbon sink implementation entities by formulating specific subsidy policies, tax incentives, and research and development funding support. At the same time, promoting the research and application of green ICT technologies and facilitating the widespread use of low-carbon technologies will provide strong technical support for carbon sink projects. In addition, governments can guide and support the development of carbon markets by providing more transparent and efficient carbon credit trading mechanisms, thereby promoting the innovation and application of green technologies. Intergovernmental cooperation on policies is also crucial. By jointly developing international standards and promoting the sharing and interconnection of green technologies, the implementation of global carbon sink projects can be accelerated.

\paragraph{\bf Multidisciplinary and international collaborations}
The successful implementation of carbon sink projects requires interdisciplinary collaboration and global coordination \cite{ontl2020forest, barman2024synergizing, leal2024university}. Carbon sink is a multifaceted challenge that spans economics, environmental science, computer science, and policy studies. Economics and environmental science provide frameworks for cost-benefit analysis and impact assessments, while computer science contributes technical solutions like big data analysis and AI-driven monitoring. Policy studies ensure the sustainability and effectiveness of these efforts. Global applications of carbon sinks would be based on international cooperation, especially for large-scale projects such as forest conservation and land restoration. Collaborations among governments, international organizations, and research institutions facilitate technology exchanges, address policy and resource disparities, and support the achievement of global carbon reduction objectives.

\section{Acknowledgments}
This work was supported in part by the China Guangxi Science and Technology Plan Project under grant AD23026096, Chile CONICYT FONDECYT Regular under grant 1181809, and Chile CONICYT FONDEF under grant ID16I10466.

\bibliographystyle{elsarticle-num}
\balance

\bibliography{fixed_egbib_org}

~~~\\
~~~\\

\section*{Biographies}
\subsection*{Yuze Dong}
Yuze Dong is currently pursuing his Master of Science degree in the School of Artificial Intelligence, Guilin University of Electronic Technology, China. His current research interests include artificial intelligence and applications, and green information technologies.

\subsection*{Jinsong Wu}
Jinsong Wu received  Ph.D. degree from Department of Electrical and Computer Engineering, Queen's University, Kingston, Canada in 2006. He received the 2020 IEEE Green Communications and Computing Technical Committee Distinguished Technical Achievement Recognition Award for his outstanding technical leadership and achievement in green wireless communications and networking. He was the leading Editor and co-author of the comprehensive book, entitled “Green Communications: Theoretical Fundamentals, Algorithms, and Applications”, published by CRC Press in September 2012. He won 2017,2019,2021 IEEE System Journal Best Paper Awards. He won the 2018 IEEE Green Communications and Computing Technical Committee Best Magazine Paper Award. He was the founder and founding Chair (2011-2017) of IEEE Technical Committee on Green Communications and Computing (TCGCC).
\end{document}